\documentclass[fleqn,usenatbib]{mnras}

\usepackage{newtxtext,newtxmath}

\usepackage[T1]{fontenc}
\usepackage{ae,aecompl}

\usepackage{eso-pic}

\AddToShipoutPictureBG*{%
  \AtPageUpperLeft{%
    \hspace{0.75\paperwidth}%
    \raisebox{-4.5\baselineskip}{%
      \makebox[0pt][l]{\textnormal{FERMILAB-PUB-18-106}}
}}}%

\usepackage{graphicx}	
\usepackage{amsmath}	
\usepackage{amssymb}	
\usepackage[shortlabels]{enumitem}

\usepackage{multicol}

\title[Transfer learning for galaxy morphology]{Transfer learning for galaxy morphology from one survey to another}

\author[ Dom\'inguez S\'anchez et al.]{
\parbox{\textwidth}{
\Large
H.~Dom\'{i}nguez S{\'a}nchez$^1$\thanks{Corresponding author: \texttt{\rm \texttt{helenado@sas.upenn.edu}}}, 
  M.~Huertas-Company$^{1, 2, 3, 4, 5}$,
  M.~Bernardi$^{1}$, 
  S. Kaviraj$^{6}$, 
  J.L. Fischer$^{1}$,
  T.~M.~C.~Abbott$^{7}$,    
  F.~B.~Abdalla$^{8, 9}$,            
  J.~Annis$^{10}$,          
  S.~Avila$^{11}$,         
  D.~Brooks$^{8}$,         
  E.~Buckley-Geer$^{10}$, 
  A.~Carnero~Rosell$^{12, 13}$, 
  M.~Carrasco~Kind$^{14, 15}$,    
  J.~Carretero$^{16}$,   
  C.~E.~Cunha$^{17}$,      
  C.~B.~D'Andrea$^{1}$,  
  L.~N.~da Costa$^{12, 13}$,    
  C.~Davis$^{17}$,          
  J.~De~Vicente$^{18}$,  
  P.~Doel$^{8}$,                
  A.~E.~Evrard$^{19, 20}$,              
  P.~Fosalba$^{21, 22}$,                
  J.~Frieman$^{10, 23}$,         
  J.~Garc\'ia-Bellido$^{24}$,
  E.~Gaztanaga$^{21, 22}$,             
  D.~W.~Gerdes$^{21, 22}$,            
  D.~Gruen$^{17, 25}$,                    
  R.~A.~Gruendl$^{14, 15}$,           
  J.~Gschwend$^{12, 13}$,             
  G.~Gutierrez$^{10}$,          
  W.~G.~Hartley$^{8, 26}$,      
  D.~L.~Hollowood$^{27}$,   
  K.~Honscheid$^{28, 29}$,            
  B.~Hoyle$^{30, 31}$,                    
  D.~J.~James$^{32}$,        
  K.~Kuehn$^{33}$,         
  N.~Kuropatkin$^{10}$,   
  O.~Lahav$^{8}$,          
  M.~A.~G.~Maia$^{12, 13}$,         
  M.~March$^{1}$,           
  P.~Melchior$^{34}$,        
  F.~Menanteau$^{14, 15}$,                    
  R.~Miquel$^{16, 35}$,         
  B.~Nord$^{10}$,            
  A.~A.~Plazas$^{36}$,       
  E.~Sanchez$^{18}$,       
  V.~Scarpine$^{10}$,       
  R.~Schindler$^{25}$,      
  M.~Schubnell$^{20}$,       
  M.~Smith$^{37}$,           
  R.~C.~Smith$^{7}$,       
  M.~Soares-Santos$^{38}$,         
  F.~Sobreira$^{39, 12}$,       
  E.~Suchyta$^{40}$,         
  M.~E.~C.~Swanson$^{15}$,   
  G.~Tarle$^{20}$,          
  D.~Thomas$^{11}$,          
  A.~R.~Walker$^{7}$,       
  and J.~Zuntz$^{41}$ }
  \vspace{0.4cm}\\~\\
\parbox{\textwidth}{\centering \textsc{\Large  } \\ \centering \textit{Author affiliations are listed at the end of this paper} }
\vspace{-1cm}
}

\pubyear{2018}

\begin{document}
\label{firstpage}
\pagerange{\pageref{firstpage}--\pageref{lastpage}}
\maketitle

\begin{abstract}

Deep Learning (DL) algorithms for morphological classification of galaxies have proven very successful,  mimicking (or even improving) visual classifications. However, these algorithms rely on large training samples of labeled galaxies (typically thousands of them).  A key question for using DL classifications in future Big Data surveys  is how much of the knowledge acquired from an existing survey can be exported to a new dataset, i.e. if the features learned by the machines are meaningful for different data. We test the performance of DL models, trained with Sloan Digital Sky Survey (SDSS)  data, on Dark Energy survey (DES) using images for a sample of ~5000 galaxies with a similar redshift distribution to SDSS.  Applying the models directly to DES data provides a reasonable global accuracy ($\sim$ 90\%), but small completeness and purity values. A fast domain adaptation step, consisting in a further training with a small DES sample of galaxies ($\sim$500-300), is enough for obtaining  an accuracy  > 95\%  and  a significant improvement in the completeness and purity values. This demonstrates that, once trained with a particular dataset, machines can quickly adapt to new instrument characteristics (e.g.,  PSF, seeing, depth), reducing by almost one order of magnitude the necessary training sample for morphological classification. Redshift evolution effects or significant depth differences are not taken into account in this study.

\end{abstract}

\begin{keywords}
  galaxies: structure  -- methods: observational -- surveys
\end{keywords}



\section{Introduction}

Astronomy is entering the Big Data era. We are experiencing a revolution in terms of available data thanks to surveys such as  COSMOS \citep{Scoville2007}, SDSS \citep{Eisenstein2011}, DEEP2 \citep{Newman2013},  DES \citep{DES2016}, etc. The close future is even brighter with missions like EUCLID \citep{Racca2016} or LSST \citep{LSST2017}, offering photometric, quasi-spectroscopic data of millions/billions of galaxies. 

One key measurement severely affected by this Big Data transition is galaxy morphology estimated from images. Galaxies exhibit a great variety of shapes and their morphology is closely related to their stellar content. In addition,  the light-profiles provide information about their mass-assembly, interactions, accretion, quenching processes or feedback (e.g. \citealt{Conselice2003, Kaviraj2014, Bournaud2014, Belfiore2015, Dubois2016}). It is therefore crucial to have accurate morphological classifications for large samples of galaxies.

	Galaxy morphological catalogues have been usually based on visual classifications. Unfortunately, visual classification is an incredible time consuming task. The size of present and future Big Data surveys, containing millions of galaxies, make this approach a  near impossible task. One beautiful solution to this problem was the Galaxy Zoo project \citep{Lintott2011}, which involved more than 100k volunteer citizens  to morphologically classify the full SDSS sample and has now been extended to other higher redshifts and surveys (e.g. CANDELS survey, \citealt{Simmons2016}; DECaLS survey). However, with the next generation of surveys, we are reaching the limit of applicability of these approaches. It is estimated that about a hundred years would be needed to classify all data from the EUCLID mission with a Galaxy Zoo-like approach, unless the number of people involved  is significantly increased. A question naturally arises: can human classifiers be replaced by algorithms? 
    
 Automated classifications using a set of parameters that correlate with morphologies, e.g.  CAS-methods (Concentration-Asymmetry-Smoothness, \citealt{Conselice2003}) or Principal Component Analysis (\citealt{Lahav1995,Lahav1996,Banerji2010}, and references therein)  have  been attempted. However, the parameter extraction also requires large amounts of time. DL algorithms where, in contrast to classic machine learning algorithms, no image pre-processing is needed, have come to the rescue for image analysis of large data surveys.  The use of  convolutional neural networks (CNNs) to learn and extract the most meaningful features at pixel level have been shown to produce excellent results for pattern recognition in complex problems and are widely used by many technology giants such as \textit{Google}. CNNs have demonstrated their success for morphological classification of galaxies in The Galaxy Challenge\footnote{https://www.kaggle.com/c/galaxy-zoo-the-galaxy-challenge}, a Kaggle competition for reproducing the Galaxy Zoo 2, where the top three algorithms used CNNs (e.g.  \citealt{Dieleman2015}). At higher redshifts, \citealt{Huertas2015} also showed that CNNs represent a major improvement with respect to CAS-based methods.

 In a companion paper, \citet[DS18 hereafter]{DS2018}, we combine the best existing visual classification catalogues with DL algorithms to provide the largest (670,000 galaxies from DR7-SDSS survey) and most accurate morphological catalogue  to date. The catalogue includes two flavours: T-Type, related to the Hubble sequence, and Galaxy Zoo 2 classification scheme. One of the main improvements with respect to previous works \citep{Dieleman2015}, is that only  galaxies with robust classifications (large agreement between Galaxy Zoo classifiers)  are used for training each task. This helps the models to detect the relevant features for each question and a smaller training sample is required for the models to converge.

In spite of this improvement on the training approach, these algorithms still rely on large training sets (around 5000-10000 galaxies, depending on the classification task). A key question, in view of using DL based algorithms to assess the morphologies of galaxies in future Big Data surveys, is therefore how much of the knowledge acquired from an existing survey can be exported to a new dataset, i.e., can the features learned by an unsupervised process on a given dataset be \textit{transferred} to a new dataset with different properties? And - if not - what is the cost of updating those features (in terms of new objects to be classified from the new dataset)?

This process, usually referred to as \textit{transfer learning} or \textit{fine-tuning} in the literature, is becoming popular for general image recognition (e.g.  \citealt{Bengio2012,Yosinski2014,Tajbakhsh2016}) and several recent works explore the optimal  strategy to \textit{transfer  knowledge} (e.g., \citealt{Shermin2018,Guo2018,Kornblith2018} and references therein). However,  \textit{transfer learning} using astronomical data has not been yet fully explored.

Some preliminary tests have  been performed by our team to assess the performance of DL algorithms, trained with simulated data, on real data. In a recent paper \citep{Tuccillo2017} we show that a DL machine trained on one-component S\'ersic galaxy simulations (with real HST/CANDELS F160W PSF and noise) can accurately recover parametric measurements of real HST galaxies with at least the same quality as GALFIT \citep{Peng2002}, but several orders of magnitude faster. It shows indications that DL is able to transition from simplistic simulations to real data without seriously impacting the results. 

 In a recent paper, \citet{Ackermann2018} investigate \textit{transfer learning} for galaxy merger detection  by retraining CNNs first trained on pictures of everyday objects (i.e., ImageNet data set, \citealt{Deng2009}).
In this work we study  \textit{transfer learning}  for morphological classification of galaxies between different astronomical surveys.  To that end, we take advantage of the DL models trained with  SDSS data  to test their performance when applied to DES  survey, with and without training on DES images. This is, to the best of our knowledge, the first work addressing the the ability of DL models  to \textit{transfer knowledge} for different datasets. In a recent work, \citet{PC2018}  provide a  morphological catalogue of CLASH \citep{Postman2012}  galaxies by \textit{fine-tuning} a CNN pre-trained on CANDELS survey \citep{Grogin2011ApJS..197...35G}. They confirm the result presented in this paper: that  \textit{transfer learning} reduces the number of labeled images needed for training. 

The paper is organised as follows: In Section \ref{sect:data} we describe the SDSS-based DL models,  DES images  and morphological catalogue used in this work; in Section \ref{sect:methods} we explain our methodology,  in Section \ref{sect:results}  we discuss the results and in Section \ref{sect:conclusions} we summarise our conclusions.

\section{Data}
\label{sect:data}

\begin{figure*}
\setlength{\columnsep}{20pt}
\begin{multicols}{2}
    \includegraphics[width=\linewidth]{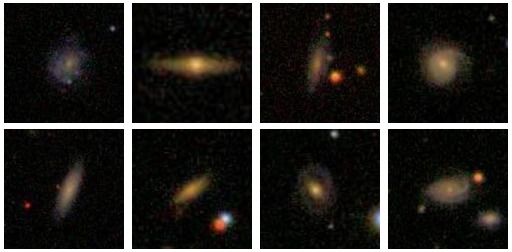}\par 
    \includegraphics[width=\linewidth]{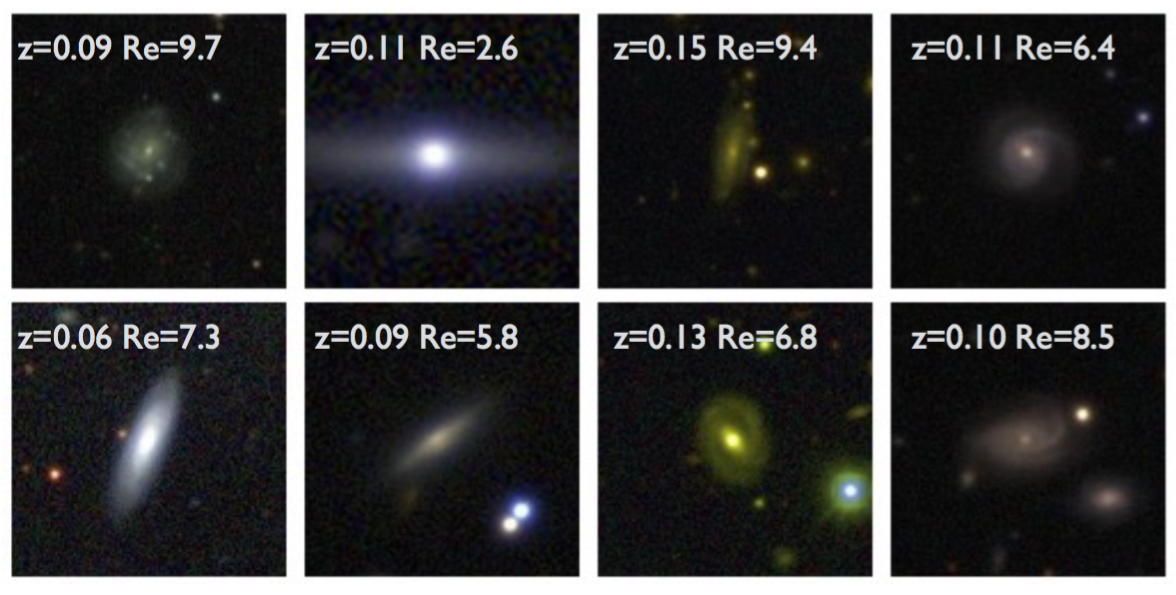}\par 
    \end{multicols}
\caption{ Examples of  6 galaxies observed by SDSS-DR7 (left panels) and DES survey (right panels). The cutouts are zoomed in  to 1/2 of the  size of the images used for training the models. They  have a variable angular size  of approximately 5$\times$R$_{90}$, where R$_{90}$ is the Petrosian radius of each galaxy (shown in each cutout - in arcsec -,  as well as their redshift). The galaxies are randomly selected from the  common sample of the two surveys, with the only requirement of having high probability of being disk, edge-on or barred galaxies. The better quality of DES images reveals with higher detail some galaxy features, such as bulge component or spiral arms.} \label{fig:cutouts}
\end{figure*}

In this paper we test the performance of DL  models, trained with SDSS-DR7 data \citep{Abazajian2009}, on DES images. The  morphological classification of DES galaxies comes from the  DECaLS - Galaxy Zoo catalogue. In this section we describe the SDSS DL models, DES images and the morphological catalogue used throughout the paper.

\subsection{Deep Learning models trained with SDSS-DR7 data}
In DS18 we  morphologically classify $\sim$670,000 SDSS-DR7 galaxies with automated DL  algorithms. The galaxies  correspond to the sample  for which \citet{Meert2015, Meert2016} provide accurate photometric reductions. Reader can refer to DS18 for a detailed explanation on the data and methodology but, in short, we use  two visual classification catalogues, Galaxy Zoo  2 (GZ2 hereafter,  \citealt{Willett2013}) and \citet{Nair2010}, for training CNNs with color  SDSS-DR7 images. We obtain T-Types and a series of GZ2 type questions (disk/features, edge-on galaxies, bar signature,  bulge prominence, roundness and mergers) for a sample of galaxies with r-band Petrosian magnitude limits $14 \le m_r \le 17.77$ mag and z < 0.25. The SDSS images are the standard cutouts downloaded from the SDSS DR7 server\footnote{http://casjobs.sdss.org/ImgCutoutDR7}, with a resolution of  0.396~\arcsec/pixel.

\subsection{Image data: Dark Energy Survey }
The images used to test how DL models can adapt to new surveys characteristics come from the  Dark Energy Survey (DES;  \citealt{DES2016}). DES  is an international, collaborative effort designed to probe the origin of the accelerating universe and  the nature of dark energy by measuring almost the 14-billion-year history of cosmic expansion with high precision.  The survey will map $\sim$ 300 million galaxies. This huge number demands to find automated methods for morphological classification of galaxies.  

DES  is a photometric survey utilizing the Dark Energy Camera (DECam; \citealt{Flaugher2015}) on the Blanco-4m  telescope  at  Cerro  Tololo  Inter-American Observatory (CTIO) in Chile to observe $\sim$5000 deg$^2$ of the  southern  sky  in  five  broad-band  filters, \textit{g}, \textit{r}, \textit{i},  \textit{z} and  \textit{Y}   ($\sim$ 400 nm to  $\sim$1060 nm)  with a resolution of  0.263~\arcsec/pixel. The magnitude limits and median PSF FWHM for the first year data release  (Y1A1 GOLD) are 23.4, 23.2, 22.5, 21.8, 20.1 mag and 1.25, 1.07, 0.97, 0.89, 1.07 arcsec, respectively (from \textit{g} to \textit{Y}, see \citealt{Drlica-Wagner2017} for a detailed description of the survey). In this work we use standard DES cutouts from the internal Y1A1 data release.

\subsection{Morphological catalogue: Dark Energy Camera Legacy Survey }
\label{sect:catalogue}
Unfortunately, there is no morphological classification available for DES galaxies to date. Instead, we take advantage of the Galaxy Zoo Dark Energy Camera Legacy Survey (DECaLS) morphological catalogue to assign a  classification for DES galaxies. This is necessary for quantifying the performance of the  DL models, as well as for labeling the training sample in the fine-tunning or domain adaptation step (see section \ref{sect:methods}).
The DECaLS survey \citep{DECALS2018} is observed with the same camera as the DES survey and with a similar depth (\textit{g}=24.0,  \textit{r}=23.4, \textit{z}=22.5 mag at 5$\sigma$ level), and so (average) observing conditions are very similar to the DES ones. The  DECaLS Galaxy Zoo catalogue (private communication) contains morphological classifications for $\sim$ 32,000 objects up to z $\sim$ 0.15. The redshift range and most of the classification tasks are the same as for the GZ2 catalogue, which was used for training the DL models from DS18. Therefore,  it is the perfect catalogue to test the performance of the SDSS-based DL models on DES images.  The main difference of DES/DECaLS with respect to SDSS images is the use of a larger telescope and better seeing conditions, which allow to get deeper images ($\sim$ 1.5 mag)  with significantly better data quality than SDSS. This effect can be seen in Figure \ref{fig:cutouts}, where we show 6 examples of galaxies as observed by SDSS and DES.  

The DES sample used in this work are  the 4,938  galaxies  with a DECaLS - Galaxy Zoo classification  (obtained with a match of 1 arcsec separation). Note that, since our final aim will be to provide a morphological catalogue for DES, we use the DECaLS classification catalogue as the ground truth  to test  (and train) our models on DES images. Given the similarities between DES and DECaLS surveys, the Galaxy Zoo classifications  will be identical or very similar, which allows us to perform this exercise.

\section{Methodology}
\label{sect:methods}

The objective of this paper is to assess if knowledge acquired by a DL algorithm from an existing survey can be exported to a new dataset with different characteristics in terms of depth, PSF and instrumental effects.  This work aims to be a first \textit{proof of concept} and not a full  morphological classification catalogue.
The redshift distribution of the DES galaxies used in this work is very similar to the SDSS (see \ref{sect:catalogue}), so no evolution effects are included: we are only changing the instrument and survey depth and spatial resolution. We leave for a forthcoming work a thoughtful study on the brightness and redshift effect on the models performance. 

We focus our analysis on the binary questions from the GZ2 scheme, since they are the  easiest  to evaluate. We note that there is one model per question. The three classification tasks that we evaluate are:

{\bf Q1}: Galaxies with disks/features versus smooth galaxies. We consider as positive examples galaxies with disk or features (Y=1 in our input label matrix). {\bf Q2}: Edge-on galaxies versus face-on galaxies. Edge-on galaxies are considered positive cases. {\bf Q3}: Galaxies with bar signature versus galaxies with no bar presence. Barred galaxies are positive cases.

\subsection{Deep Learning architecture}

The methodology used in this paper (in terms of training sample selection, model input and DL model architecture) is exactly the same as in DS18, where the reader can find a detailed explanation about the procedure.  In this study we do not aim to maximize absolute model performance,  but rather to study knowledge transfer on a well-known architecture.  To facilitate the reader, in Table \ref{tab:CNN} we summarise the DL model architecture, which consists on four convolutional layers (with ReLU activation, Max Pooling and dropout)  and one fully connected layer (also referred to as the dense layer). The total number of free parameters is 2,602,849 (see also Figure 1 in DS18).

\begin{table}
\centering
\begin{tabular}{|l| c|c|}

\hline
 Layer type   &  Output shape & Num. parameters \\
\hline
\hline

Conv2D  (6x6)   &   (32, 69, 69)    &    3488   \\
Dropout  (0.5)  &   (32, 69, 69)    &   0          \\
\hline
Conv2D    (5x5)      &    (64, 69, 69)    &    51264     \\
MaxPooling  &  (64, 34, 34)    &    0         \\
Dropout   (0.25)      &    (64, 34, 34)     &    0         \\
\hline
Conv2D  (2x2)  &        (128, 34, 34)     &  32896     \\
MaxPooling  & (128, 17, 17)   &    0          \\
Dropout  (0.25)  &          (128, 17, 17)   &     0     \\
\hline
Conv2D    (3x3)    &        (128, 17, 17)      &     147584  \\   
Dropout  (0.25)     &       (128, 17, 17)    &        0     \\     
Flatten       &      (36992)      &           0          \\
\hline
Dense         &         (64)         &           2367552   \\   
Dropout   (0.5)  &          (64)       &             0          \\
Dense         &     (1)          &           65         \\
\hline
\hline
Total num. parameters & &   2602849 \\
\hline
\end{tabular}
\caption{DL model architecture. It consists of 4 convolution layers with different filter sizes (6, 5, 2 and 3, as shown in brackets) and one fully connected layer, also referred to as the dense layer. Dropout is performed after each convolutional layer (the reduction factor is shown in brakets) and MaxPooling is used after the second and third layers. The output shape and the number of free parameters in each layer are also shown.}
\label{tab:CNN}
\end{table}

To keep the methodology as similar as possible to DS18,  the input for the models are the same as in DS18, i.e. 424$\times$424 pixel  size images (from DES in this case), which are down-sampled into (69, 69, 3) RGB matrices, with each number representing the flux per pixel at each filter (\textit{g, r, i}).  The flux values are normalized to the maximum value in each filter for each galaxy. The angular size  of the images is variable, approximately   10$\times$R$_{90}$, where R$_{90}$ is the Petrosian radius of each galaxy (from SDSS). 

\subsection{Training and transfer learning}
\label{sect:TK}

In order to  assess how much knowledge from one survey can be exported to another, we carry out four experiments:

\begin{enumerate}[leftmargin=0.5cm,label={\alph*)}]
	\item  Apply the models trained on SDSS data  directly to DES images, without any further training or \textit{fine-tunning} on DES data.
    \item  Load the weights trained  on SDSS data and  \textit{fine-tune} them by training the models with a small DES sample (300-500 galaxies). The training is performed for all the layers in the DL model.
	\item  Same as (b) but \textit{freezing} all the layers (i.e., fixing the weights learned by SDSS) except the fully connected layer.
    \item Training the models from scratch using a DES training sample with the same size as in (b) and (c). 
\end{enumerate}

We compare these experiments with the results presented in DS18  for  models trained and tested on SDSS data. Note that in this work we focus on  knowledge transfer between different datasets, not between different tasks. This means that, for experiments (a) to (c), we use the SDSS models trained for each particular task.

 For test  (a), the algorithm applies the weights learned by the SDSS models and returns a probability value for each task. For tests  (b) to (d) the training procedure is identical to the one used in DS18.  We train the models in binary mode. Data augmentation (as explained in DS18) is applied to the DES images to help avoiding overfitting. Balanced weights are used for Q2 and Q3 due to the uneven proportion of positive and negative examples for this two classes. We only use in the training DES galaxies with a robust classification, i.e. galaxies with a large agreement - $a(p)$ -  between Galaxy Zoo classifiers (roughly corresponding to P > 0.7 in one of the two answers) and with at least 5 votes.  [Reader can refer to DS18 for a description of the agreement parameter, $a(p)$.] This methodology has demonstrated to be a more efficient way to train the models,  but it strongly limits the statistics of our train and test samples. For example, only 624 out of 4938  galaxies  ($\sim$ 13\%)  have $P_{edge-on}$ > 0.7 and at least 5 votes. This number is even smaller (103, $\sim$ 2\%) for the barred galaxies.  
 
 We test the fine-tuned models  on a sample of DES galaxies not used for training. Although this limits the statistics, specially in the case of Q3 (bar signature), it is important to properly evaluate the models. Since we  need at least 300 galaxies for training Q3 (and the training sample should include a reasonable number of positive cases), we only have 9 barred galaxies left  for testing our models (see Table \ref{tab:TPR-P-Acc}). The code used in this work is publicly available at \texttt{https://github.com/HelenaDominguez/DeepLearning}. 

\section{Results and discussion}
\label{sect:results}

\begin{figure*}
\setlength{\columnsep}{1pt}
\begin{multicols}{3}
    \includegraphics[width=\linewidth]{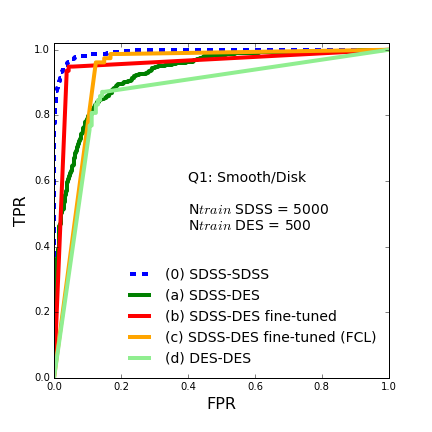}\par 
    \includegraphics[width=\linewidth]{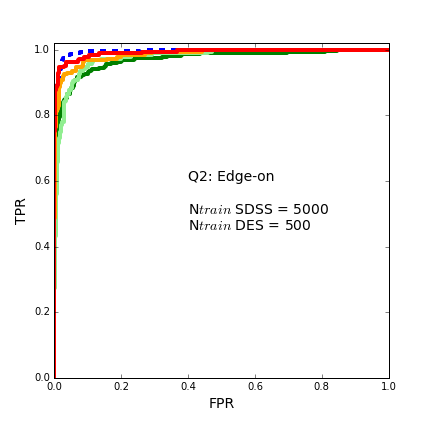}\par 
     \includegraphics[width=\linewidth]{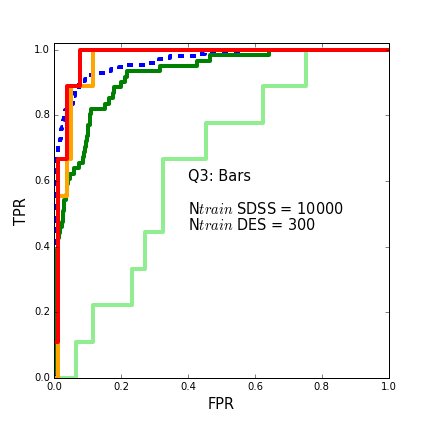}\par 
    \end{multicols}
\caption{ True positive rate (TPR, i.e., fraction of well classified positive cases) vs. false positive rate (FPR, i.e., fraction of wrongly classified positive cases)  for different P$_{th}$ values for the three classification task studied in this work, as stated in the legend. We show the performance of the DL models for the four experiments explained in section \ref{sect:TK} (labeled (a) to (d), color-coded as shown in the legend), as well as the results of the models trained with SDSS galaxies and applied to SDSS images (blue dashed line). The number of galaxies used in the training for each question for the SDSS and DES samples are shown in the legend. The knowledge transfer from SDSS plus \textit{fine-tunning} provides results comparable to the SDSS-SDSS models, but the training sample size can be reduced at least one order of magnitude. The `apparent' better performance of the \textit{fine-tuned}  DES models with respect to the SDSS-SDSS one for Q3 is caused by the small size of the barred test sample  (see Table \ref{tab:TPR-P-Acc}). } \label{fig:ROC}
\end{figure*}

\begin{table*}
\centering
\begin{tabular}{l|c| c| c|c| c|c| c|c|}
Question  &   & Experiment  &   N$train$       &   N$test$ & N$pos$   & TPR    &  Prec.  &  Acc.    \\
\hline

                                   & (0)  &  SDSS-SDSS                                  &   5000   & 3370  & 674 &  0.93  &  0.91  &  0.97      \\
         Q1                    & (a)   &  SDSS-DES                                    &  0           &  2409 & 797  & 0.48   &  0.92   &  0.81    \\
   Smooth/Disk    & (b)   &	 SDSS-DES fine-tuned              &  500     & 238     &78      & 0.95   &  0.91   &  0.95        \\
                                   & (c) &	 SDSS-DES fine-tuned (FCL)  &  500      & 238     &78      & 0.96   &  0.78   &  0.90        \\
                                   & (d) &	 DES-DES                                       &  500     & 238     &78       & 0.81  &  0.77   &  0.85        \\
\hline

                         & (0)  &  SDSS-SDSS                                   & 5000  & 2687 & 396    &  0.98 &  0.80  &  0.96      \\
         Q2           & (a)  &   SDSS-DES                                     & 0         & 2851 &  536    & 0.91   &  0.76   &  0.93    \\
      Edge-on	& (b) &	  SDSS-DES fine-tuned              & 500     & 738 & 187       & 0.96   &  0.86   &  0.95        \\
                         & (c) 	&	  SDSS-DES fine-tuned (FCL) & 500     & 738 & 187       & 0.97   &  0.77   &  0.92   \\
                          & (d) &	 DES-DES                                       &  500    & 238  &78         & 0.97   &  0.70   &  0.89        \\
\hline
                        & (0)    &  SDSS  & 10000  & 1806 & 169 &  0.76  &  0.79  &  0.96      \\
         Q3         & (a)     &   DES  &      0   & 1768 & 61    &  0.57   &  0.35   &  0.95    \\
   Bar sign		& (b) 	&	  SDSS-DES fine-tuned   & 300  & 86 & 9 & 0.89  &  0.73   &  0.95  \\
    		            & (c)    &	  SDSS-DES fine-tuned (FCL)  & 300  & 86 & 9 & 1.0  &  0.5   &  0.89       \\
                        & (d)    &	 DES -DES  & 300  & 86 & 9 & 1.0  &  0.1   &  0.1      \\
\hline
\end{tabular}
\caption{Performance of the models according to the TPR, precision and accuracy values for the three classification tasks studied in this work. The experiment column specifies the approach used, as explained in section \ref{sect:TK}. N$train$ is the number of galaxies used for training. When N$train$=0, it means the SDSS model is directly applied to DES data. N$test$ are the number of galaxies used for testing the models (they fulfill the requirement of having a robust morphological classification, as the training sample), of which N$pos$ are the positive cases (e.g., galaxies showing disk/features for Q1).  Galaxies used for training are not included in the testing sample. This explains the scarcity of barred  galaxies used for testing the models with DES training.}
\label{tab:TPR-P-Acc}
\end{table*}

We use a standard method for testing the performance of our models: receiver operating characteristic  (ROC) curve, true positive rate (TPR, also known as recall), precision (P) and accuracy values (e.g., \citealt{Powers2011}, \citealt{Dieleman2015}, \citealt{Barchi2017}). For binary classifications,  where only two input values are possible (positive or negative cases),  the true positives (TP) are the correctly classified positive examples. One can define,  in an analogous way, true negatives, false positives, and false negatives (TN, FP, FN, respectively). The true positive rate (TPR), false positive rate (FPR),  precision (P) and accuracy  (Acc) are  expressed as:

\begin{equation}
    \begin{aligned}
{ \rm TPR=\frac{TP}{(TP+FN)};   \qquad FPR=\frac{FP}{(FP+TN)}}\\[10pt]
{ \rm P=\frac{TP}{(TP+FP)} ;  \qquad Acc=\frac{TP+TN}{Total}} \\
   \end{aligned}
\end{equation}
TPR is a  completeness proxy (how many of the true examples are recovered), precision is a contamination indicator (what fraction of the output positive cases are really positive) and accuracy is the fraction of correctly classified objects among the test sample. Since the output of the model is  a probability (ranging form 0 to 1), a  probability threshold (P$_{th}$) value must be  chosen  to separate positive and negative cases. The ROC curve represents the TPR and FPR values for different P$_{th}$.  A perfect classifier would yield a point in  upper left corner or coordinate (0,1) of the ROC space,  (i.e., no false negatives and no false positives), while a random classifier would  give a point along a diagonal line.

In Figure \ref{fig:ROC} we show the ROC curve for the three classification tasks studied in this work for  the SDSS model applied to SDSS data (0), for the SDSS model applied to DES data without any training on DES  (a), for the model \textit{fine-tuned} on a small DES sample with transfer knowledge from the SDSS model (allowing all layers to be trained (b) or freezing all layers but the fully connected layer (c)), and for the model trained with a random initialization on a small DES sample (d). 

In Table \ref{tab:TPR-P-Acc}, we show the TPR, precision and accuracy values for the same experiments. For simplicity, we only list the values obtained for P$_{th}$= 0.5 (the standard value for separating positive and negative cases).  Both the train and  test DES samples are required to have a robust classification  in the morphological catalogue (see section \ref{sect:methods}). The number of galaxies used for training and testing (and the positive cases), are also given in Table \ref{tab:TPR-P-Acc}.

 Our first main result is that, when applying the SDSS-models directly to DES images, with no training at all on DES data,  the accuracy values obtained are reasonable (> 80\%), reaching 93\% and  95\% for Q2 and Q3.  However, the accuracy can be misleading  when few positive cases are included in the test sample and it is important to consider completeness and purity of the classification.  This quantities are strongly dependent on the classification task. For example, for Q1 the precision value is very high (92\%), but the completeness is less than 50\%. On the other hand, the SDSS model recovers 91\% of the DES edge-on galaxies, but the precision value for this task is 76\%. For Q3, both the  completeness and purity values obtained with the SDSS model are  small (0.57 and 0.35, respectively).  This indicates that  bar  identification is a very sensitive task to resolution and depth,  while, on the other hand,  inclination is less dependent on the survey characteristics. 

\begin{figure}
 \includegraphics[width=\linewidth]{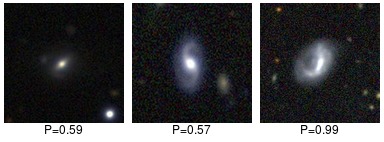}\par 
\caption{ The 3 FP cases of barred galaxies,  according to the DES-fine-tuned model (b). The numbers shown in the cutouts are the predicted $P_{bar}$ given by our model. The cutouts are zoomed in to $\sim$ 1/2 of the input to the models (approximately   5$\times$R$_{90}$). There is a clear bright central structure in all of them, which may be difficult to distinguish from a  true bar, even for non-expert human classifiers.} \label{fig:FP-bars}
\end{figure}

 The second main result is that, after a fast domain adaptation step (i.e.,  training the models with a small sample - less than 500 - of highly reliable DES galaxies), the models are able to adapt to the new data characteristics and quickly converge, providing results comparable to the ones obtained for the SDSS models applied to SDSS data (see Table \ref{tab:TPR-P-Acc} and Figure \ref{fig:ROC}).  We tested the performance of the models with DES training samples of different sizes and we found that the presented here are an optimal trade-off between models' results and training sample size.  The accuracy values are $\geq$ 0.95 for all the classification tasks. For both Q1 and Q2 the completeness reaches at least 95\% and the purity values are 91\% and 86\%, respectively. The TPR and precision values for Q3 are smaller (0.89 and 0.73, respectively), but are severely affected by the test sample statistics. In fact, the model recovers 8 out of 9 barred galaxies (TP) and there are only 3 FP cases. After visual inspection, we found that the FN case is not a real barred galaxy but a bulge dominated galaxy. On the other hand, only one of the 3 FP cases have $P_{bar}$ > 0.6 according to  model (b), and that galaxy shows a  bright central feature which could  be a distorted bar or a dust lane (see Figure \ref{fig:FP-bars}).
 
 Regarding the comparison between experiments (b) and (c), the results are slightly better for all tasks when training both the convolutional filters and the fully connected layer, rather than training  the fully connected layer alone.  Given the `simplicity' of the CNN used (only 4 convolutional layers), most of the trainable weights actually come from the fully connected layer (235232 vs  2367617 for the convolutional layer and the fully connected layer, respectively). Despite of this, the performance of the models after \textit{fine-tunning} all the layers is improved. It has been suggested in the literature (e.g., \citealt{Yosinski2014}) that the first-layer features  of deep neural networks are \textit{general}, in the sense that they can be applicable to many datasets and tasks.   The results from this work indicate that the features learned by the convolutional layers are in fact important to improve the classification.  Note that \citealt{Yosinski2014} work is based on different classification tasks using the same input images, while in this work we want to transfer knowledge between different surveys. Our results suggest that the differences arising from different data sets (i.e., the survey image characteristics) have an effect on the features learned by the CNN, and not only by the dense layer.
 
 To better understand the impact of  \textit{transfer learning} from the SDSS models, we train the models  with the same DES training sample as in the previous exercises, but now with a random initialization. As expected, the performance of the models trained from scratch is worst  than the performance of the models after \textit{fine-tunning}. This demonstrates that using a SDSS initialization leads to a better local minimum during training. However, the results are strongly dependent of the task being trained. For example, the accuracy for Q2 is 89\% and the ROC curve is comparable to (even above) the one obtained when applying the SDSS models to the DES data without training (a). On the other hand,  a model trained with such as small sample is unable to learn and separate the features related to the presence of a bar,  as can be seen from the ROC curve shape and Table \ref{tab:TPR-P-Acc}. This reveals that CNNs efficiency is  related to the difficulty level of the classification task being trained (identifying edge-on galaxies is a much easier exercise than detecting bars).
 
 Another interesting point is the fact that the models trained with a small DES sample provide similar results to the SDSS models applied to DES data without \textit{fine-tunning} (except for Q3, as previously discussed). It suggests that \textit{transfer learning} is equivalent  to a small training step. Note that, for Q1, the area below the ROC curve of model (a) - dark green- is larger (i.e., better performance) than the one for model (d) -light green-,  although the accuracy and TRP values are smaller. This is because the values  in Table \ref{tab:TPR-P-Acc} are given for P$_{th}$=0.5, while the optimal performance for model (a) is obtained  by setting  P$_{th}$=0.1. This means that the knowledge transferred  for different datasets needs to be \textit{recalibrated}.

\section{Conclusions}
\label{sect:conclusions}

In this paper we demonstrate  that deep-nets can transfer knowledge from one survey to another and quickly adapt to new domains and data characteristics such as depth, PSF and instrumental effects. The combination of  \textit{transfer learning} and \textit{fine-tunning} boosts the models performance and allows for a significant reduction of the training sample size.

The fact that the training sample (and therefore the a priori labeled galaxies) can be reduced by an order of magnitude, once the models are trained with a different dataset, is a major discovery in order to apply DL models to future  surveys, such as EUCLID or LSST. It means that we will be able to recycle models from previous surveys (within the same redshift distribution), preventing from the huge effort of visually classifying a large sample of galaxies from that particular survey. 

It is beyond the scope of this paper to test the effect of the models on more complicated aspects of galaxy surveys, such as redshift evolution. We leave for a forthcoming work this  mandatory step to release a reliable morphological catalogue, which will certainly be an add-on value to the DES.  Also, a major advance of  extremely deep future surveys will be the detection of features which are invisible in surveys such as SDSS or DES  (e.g., tidal features and debris). Machines trained on shallower data  are unlikely to produce robust results on very deep images. We plan to carry out a thorough study to this respect using  cosmological hydro-dynamical simulations such as  Horizon-AGN \citep{Kaviraj2017}  in a future work.

\section*{Acknowledgements}

The authors thank the anonymous referee for useful comments which helped to improve the content of the paper. This work was supported in part by NSF AST-1816330. Funding for the DES Projects has been provided by the DOE and NSF(USA), MEC/MICINN/MINECO(Spain), STFC(UK), HEFCE(UK). NCSA(UIUC), KICP(U. Chicago), CCAPP(Ohio State), 
MIFPA(Texas A\&M), CNPQ, FAPERJ, FINEP (Brazil), DFG(Germany) and the Collaborating Institutions in the Dark Energy Survey. The Collaborating Institutions are Argonne Lab, UC Santa Cruz, University of Cambridge, CIEMAT-Madrid, University of Chicago, University College London, DES-Brazil Consortium, University of Edinburgh, ETH Z{\"u}rich, Fermilab, University of Illinois, ICE (IEEC-CSIC), IFAE Barcelona, Lawrence Berkeley Lab, 
LMU M{\"u}nchen and the associated Excellence Cluster Universe, University of Michigan, NOAO, University of Nottingham, Ohio State University, University of 
Pennsylvania, University of Portsmouth, SLAC National Lab, Stanford University, University of Sussex, Texas A\&M University, and the OzDES Membership Consortium. Based in part on observations at Cerro Tololo Inter-American Observatory, National Optical Astronomy Observatory, which is operated by the Association of 
Universities for Research in Astronomy (AURA) under a cooperative agreement with the National Science Foundation. The DES Data Management System is supported by the NSF under Grant Numbers AST-1138766 and AST-1536171. 
The DES participants from Spanish institutions are partially supported by MINECO under grants AYA2015-71825, ESP2015-66861, FPA2015-68048, SEV-2016-0588, SEV-2016-0597, and MDM-2015-0509, some of which include ERDF funds from the European Union. IFAE is partially funded by the CERCA program of the Generalitat de Catalunya.
Research leading to these results has received funding from the European Research
Council under the European Union's Seventh Framework Program (FP7/2007-2013) including ERC grant agreements 240672, 291329, and 306478. We  acknowledge support from the Australian Research Council Centre of Excellence for All-sky Astrophysics (CAASTRO), through project number CE110001020, and the Brazilian Instituto Nacional de Ci\^encia
e Tecnologia (INCT) e-Universe (CNPq grant 465376/2014-2). This manuscript has been authored by Fermi Research Alliance, LLC under Contract No. DE-AC02-07CH11359 with the U.S. Department of Energy, Office of Science, Office of High Energy Physics. The United States Government retains and the publisher, by accepting the article for publication, acknowledges that the United States Government retains a non-exclusive, paid-up, irrevocable, world-wide license to publish or reproduce the published form of this manuscript, or allow others to do so, for United States Government purposes.





\bibliographystyle{mnras}
\bibliography{sample} 





\section*{Affiliations}
$^{1}$ Department of Physics and Astronomy, University of Pennsylvania, Philadelphia, PA 19104, USA\\
$^{2}$ LERMA, Observatoire de Paris, PSL Research University, CNRS, Sorbonne Universit\'es, UPMC Univ. Paris 06,
F-75014 Paris, France\\
$^{3}$ University of Paris Denis Diderot, University of Paris Sorbonne Cit\'e (PSC), 75205 Paris Cedex 13, France\\
$^{4}$
Instituto de Astrof\'isica de Canarias, E-38200 La Laguna, Tenerife, Spain\\
$^{5}$
Departamento de Astrof\'isica, Universidad de La Laguna, E-38206 La Laguna, Tenerife, Spain\\
$^{6}$ Centre for Astrophysics Research, University of Hertfordshire, College Lane, Hatfield, Herts AL10 9AB, UK \\
$^{7}$ Cerro Tololo Inter-American Observatory, National Optical Astronomy Observatory, Casilla 603, La Serena, Chile\\                                 
$^{8}$ Department of Physics \& Astronomy, University College London, Gower Street, London, WC1E 6BT, UK\\                                             
$^{9}$ Department of Physics and Electronics, Rhodes University, PO Box 94, Grahamstown, 6140, South Africa\\                                           
$^{10}$ Fermi National Accelerator Laboratory, P. O. Box 500, Batavia, IL 60510, USA\\                                                                   
$^{11}$ Institute of Cosmology \& Gravitation, University of Portsmouth, Portsmouth, PO1 3FX, UK\\                                                                                                                                                                        
$^{12}$ Laborat\'orio Interinstitucional de e-Astronomia - LIneA, Rua Gal. Jos\'e Cristino 77, Rio de Janeiro, RJ - 20921-400, Brazil\\              
$^{13}$ Observat\'orio Nacional, Rua Gal. Jos\'e Cristino 77, Rio de Janeiro, RJ - 20921-400, Brazil\\                                               
$^{14}$ Department of Astronomy, University of Illinois at Urbana-Champaign, 1002 W. Green Street, Urbana, IL 61801, USA\\                               
$^{15}$ National Center for Supercomputing Applications, 1205 West Clark St., Urbana, IL 61801, USA\\                                                    
$^{16}$ Institut de F\'{\i}sica d'Altes Energies (IFAE), The Barcelona Institute of Science and Technology, Campus UAB, 08193 Bellaterra (Barcelona) Spain \\
$^{17}$ Kavli Institute for Particle Astrophysics \& Cosmology, P. O. Box 2450, Stanford University, Stanford, CA 94305, USA\\                                                                                                                                                      
$^{18}$ Centro de Investigaciones Energ\'eticas, Medioambientales y Tecnol\'ogicas (CIEMAT), Madrid, Spain\\                                                                                      
$^{19}$ Department of Astronomy, University of Michigan, Ann Arbor, MI 48109, USA\\                                                                      
$^{20}$ Department of Physics, University of Michigan, Ann Arbor, MI 48109, USA\\                                                                        
$^{21}$ Institut d\'Estudis Espacials de Catalunya (IEEC), 08193 Barcelona, Spain\\                                                                      
$^{22}$ Institute of Space Sciences (ICE, CSIC),  Campus UAB, Carrer de Can Magrans, s/n,  08193 Barcelona, Spain\\                                                                                                         
$^{23}$ Kavli Institute for Cosmological Physics, University of Chicago, Chicago, IL 60637, USA\\                                                        
$^{24}$ Instituto de Fisica Teorica UAM/CSIC, Universidad Autonoma de Madrid, 28049 Madrid, Spain\\                                                                                                                                                                                                                                                               
$^{25}$ SLAC National Accelerator Laboratory, Menlo Park, CA 94025, USA\\                                                                                                                                                                                                                                                                                                                                  
$^{26}$ Department of Physics, ETH Zurich, Wolfgang-Pauli-Strasse 16, CH-8093 Zurich, Switzerland\\                                                      
$^{27}$ Santa Cruz Institute for Particle Physics, Santa Cruz, CA 95064, USA\\                                                                           
$^{28}$ Center for Cosmology and Astro-Particle Physics, The Ohio State University, Columbus, OH 43210, USA\\                                            
$^{29}$ Department of Physics, The Ohio State University, Columbus, OH 43210, USA\\                                                                      
$^{30}$ Max Planck Institute for Extraterrestrial Physics, Giessenbachstrasse, 85748 Garching, Germany\\                                                 
$^{31}$ Universit\"ats-Sternwarte, Fakult\"at f\"ur Physik, Ludwig-Maximilians Universit\"at M\"unchen, Scheinerstr. 1, 81679 M\"unchen, Germany\\
$^{32}$ Harvard-Smithsonian Center for Astrophysics, Cambridge, MA 02138, USA\\                                                                          
$^{33}$ Australian Astronomical Observatory, North Ryde, NSW 2113, Australia\\                                                                                                                                                                                                                                                                                          
$^{34}$ Department of Astrophysical Sciences, Princeton University, Peyton Hall, Princeton, NJ 08544, USA\\                                                                                                                                
$^{35}$ Instituci\'o Catalana de Recerca i Estudis  Avan{c}cats, E-08010 Barcelona, Spain\\                                                           
$^{36}$ Jet Propulsion Laboratory, California Institute of Technology, 4800 Oak Grove Dr., Pasadena, CA 91109, USA\\                                                                                                                                                                                                                                                                                                     
$^{37}$ School of Physics and Astronomy, University of Southampton,  Southampton, SO17 1BJ, UK\\                                                                                        
$^{38}$ Brandeis University, Physics Department, 415 South Street, Waltham MA 02453\\                                                                    
$^{39}$ Instituto de F\'isica Gleb Wataghin, Universidade Estadual de Campinas, 13083-859, Campinas, SP, Brazil\\                                      
            
$^{40}$ Computer Science and Mathematics Division, Oak Ridge National Laboratory, Oak Ridge, TN 37831\\                                                                                                                                                                                                                                                                     
$^{41}$ Institute for Astronomy, University of Edinburgh, Edinburgh EH9 3HJ, UK\\

\bsp	
\label{lastpage}
\end{document}